\documentclass[sigconf]{acmart}
\usepackage{balance}
\usepackage{multicol} 
\usepackage{multirow}
\usepackage{booktabs}
\usepackage{enumitem} 
\usepackage{graphicx}
\usepackage{hyperref}
\allowdisplaybreaks
\AtBeginDocument{%
  }

\copyrightyear{2025} 
\acmYear{2025} 
\setcopyright{acmlicensed}
\acmConference[SIGIR '25]{Proceedings of the 48th
International ACM SIGIR Conference on Research and Development in
Information Retrieval}{July 13--18, 2025}{Padua, Italy}
\acmBooktitle{Proceedings of the 48th International ACM SIGIR Conference on
Research and Development in Information Retrieval (SIGIR '25), July 13--18,
2025, Padua, Italy}
\acmDOI{10.1145/3726302.3730040}
\acmISBN{979-8-4007-1592-1/2025/07}

\begin{document}


\title{MSCRS: Multi-modal Semantic Graph Prompt Learning Framework for Conversational Recommender Systems}
%

\author{Yibiao Wei}
\affiliation{%
  \institution{University of Electronic Science and Technology of China}
  \city{Chengdu}
  \state{Sichuan}
  \country{China}
}
\email{weiyibiao12138@gmail.com}

\author{Jie Zou}
\authornote{Corresponding author.}
\affiliation{%
  \institution{University of Electronic Science and Technology of China}
  \city{Chengdu}
  \state{Sichuan}
  \country{China}
}
\email{jie.zou@uestc.edu.cn}

\author{Weikang Guo}
\affiliation{%
  \institution{Southwestern University of Finance and Economics}
  \city{Chengdu}
  \state{Sichuan}
  \country{China}}
\email{guowk@swufe.edu.cn}

\author{Guoqing Wang}
\affiliation{%
  \institution{University of Electronic Science and Technology of China}
  \city{Chengdu}
  \state{Sichuan}
  \country{China}}
\email{gqwang0420@hotmail.com}

\author{Xing Xu}
\affiliation{%
  \institution{University of Electronic Science and Technology of China}
  \city{Chengdu}
  \state{Sichuan}
  \country{China}}
\email{xing.xu@uestc.edu.cn}

\author{Yang Yang}
\affiliation{%
  \institution{University of Electronic Science and Technology of China}
  \city{Chengdu}
  \state{Sichuan}
  \country{China}}
\email{yang.yang@uestc.edu.cn}


\renewcommand{\shortauthors}{Yibiao Wei et al.}

\begin{abstract}
Conversational Recommender Systems (CRSs) aim to provide personalized recommendations by interacting with users through conversations. Most existing studies of CRS focus on extracting user preferences from conversational contexts. However, due to the short and sparse nature of conversational contexts, it is difficult to fully capture user preferences by conversational contexts only. We argue that multi-modal semantic information can enrich user preference expressions from diverse dimensions (e.g., a user preference for a certain movie may stem from its magnificent visual effects and compelling storyline). In this paper, we propose a multi-modal semantic graph prompt learning framework for CRS, named MSCRS. First, we extract textual and image features of items mentioned in the conversational contexts. Second, we capture higher-order semantic associations within different semantic modalities (collaborative, textual, and image) by constructing modality-specific graph structures. Finally, we propose an innovative integration of multi-modal semantic graphs with prompt learning, harnessing the power of large language models to comprehensively explore high-dimensional semantic relationships. Experimental results demonstrate that our proposed method significantly improves accuracy in item recommendation, as well as generates more natural and contextually relevant content in response generation. Code and extended multi‑modal CRS datasets are available at https://github.com/BIAOBIAO12138/MSCRS-main.
\end{abstract}


\maketitle

\section{INTRODUCTION}
Conversational Recommender Systems (CRSs) \cite{CRS1,CRS2,CRS3}, as an emerging research direction that integrates natural language processing and recommendation technologies, aim to precisely capture users' preferences \cite{CRS4} through multi-turn conversations and thus provide personalized recommendations. Early CRS \cite{single4,single3,single2,single5,single1} primarily focused on analyzing and modeling user preferences from conversational contexts. However, due to the limited nature of conversational contexts (e.g., short and sparse), relying solely on extracting user preferences from conversations makes it challenging to achieve personalized modeling. This results in recommendations being confined to common options and failing to capture more granular user needs, thus affecting accuracy. 

\begin{figure}[tbp]
  \centering
  \includegraphics[width=0.47\textwidth]{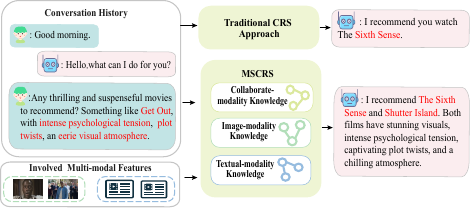}  
  \caption{Comparison between traditional CRS models and our MSCRS model.}
  \label{fig:abstract}
\end{figure}

Accordingly, to overcome this shortcoming, some approaches attempt to introduce external knowledge, including structured knowledge graph \cite{dbpedia,conceptnet}, hypergraph \cite{hycorec}, unstructured item reviews \cite{revcore,c2crs}, metadata \cite{meta}, and entities appearing in similar conversations \cite{DCRS} to enhance the user representation for improving CRS. Although these methods have made significant progress in the field of CRS, they mainly focus on a single textual modality and fail to fully utilize the rich multi-modal semantic information of items. 
As shown in Figure \ref{fig:abstract}, users' descriptions of their preferences within a single conversation are often based on their rich multi-modal experiences involving visual and textual information in reality. We argue that leveraging multi-modal semantic information is highly helpful for modeling user preferences comprehensively. First, the multi-modal features of items (e.g., posters, trailers, reviews) enrich user preference expressions from different perspectives, capturing users' multi-modal preferences that cannot be fully expressed solely through conversational contexts. Second, collaborative information, as a type of multi-modal semantic information, provides an extra perspective on user relationships (e.g., analyzing the same entities mentioned in different user conversations can uncover latent preference connections). The combination of collaborative information with multi-modal features not only enhances the modeling of explicit user preferences but also enables better identification of implicit needs.

Although we highlight the potential of multi-modal semantic information, integrating these different modalities during the conversation remains a challenge. Many existing methods enhance entity representations by incorporating external information (e.g., reviews \cite{revcore} and metadata \cite{c2crs}). However, it is well known that a semantic gap exists between conversational context and these external data, due to the inconsistencies in expression form and information structure between them, complicating the direct integration of different data sources. For instance, users expressed preferences in conversations are often subjective sentiments described through natural language, whereas reviews and metadata tend to contain more objective item characteristics and user evaluations. This makes it difficult to align semantics among different data sources, eventually affecting the performance of recommendations. While there have been some research attempts to directly fuse different data sources (e.g., contrastive learning \cite{c2crs}), the quest to align different modality data can be counterproductive because of the unique semantic associations within each modality.
To address these challenges, in this paper, we propose a novel \textbf{M}ulti-modal \textbf{S}emantic graph prompt learning framework for \textbf{CRS} (\textbf{MSCRS}). 
Specifically, first, we extract textual descriptions of items by employing large language models (LLMs) and extract images of items from an external database (i.e., IMDb\footnote{\url{https://www.imdb.com}} in this work). Afterward, we utilize pre-trained models to extract textual and image features of the items. 
Second, as direct alignment of different modality features may destroy the intra-modal semantic associations, we construct a modal-specific semantic graph for the semantic features of each modality. For collaborative modality, we extract entities (including items and item-related entities) from conversational contexts and construct a collaborative semantic graph based on the co-mention frequency of these entities. For the textual and image modalities, we construct a textual semantic graph and an image semantic graph by exploiting intra-modal feature similarity based on the extracted textual and image features. By sharing the initial embeddings of all semantic graphs, we achieve an effective fusion of the three semantic graphs. This approach avoids direct fusion and alignment of different modality features while effectively preserving the semantic relationships within each modality. Third, we propose a novel approach that integrates multi-modal semantic graphs (textual semantic graph, image semantic graph, and collaborative semantic graph) with LLMs. This integration leverages the advantages of graph neural networks (GNNs) in aggregating neighborhood information, providing topological insights to LLMs. This also enables LLMs to fully exploit high-dimensional semantic associations, guiding the selection of relevant information from textual inputs and controlling the generation process. In this way, it not only improves the performance of the recommendation task but also generates more expressive responses for the conversation task. We summarize our contributions as follows:
\begin{itemize}[leftmargin=*]
\item We propose a novel CRS model, MSCRS, which integrates multi-modal semantic information, including collaborative information and multi-modal features. To the best of our knowledge, this is the first effort to leverage both collaborative information and multi-modal item features for generation-based CRS.
\item MSCRS constructs semantic graphs based on intra-modal relations and avoids the cross-modal semantic gap via shared embedding. Additionally, it proposes a novel framework to combine multi-modal semantic graphs with prompt learning, which leverages LLMs to explore higher-order semantic associations, enabling more accurate user preference modeling and more natural response generation.
\item 
To support multi-modal CRS research, we supplemented the multi-modal features for two widely used CRS datasets. Experimental results on two widely used benchmark datasets demonstrate that our proposed MSCRS outperforms state-of-the-art baselines in both item recommendation and response generation.
\end{itemize}

\section{RELATED WORK}
\subsection{Conversational Recommendation}
As dialogue systems \cite{dialog1,dialog2,dialog3} have rapidly evolved, CRS \cite{CRS1,CRS3,CRS4} have become a thriving field of research. CRS aims to discern user preferences through multi-turn interactions and suggest items that users might find appealing. Current CRS can generally be divided into two categories: attribute-based CRS and generation-based CRS. 

Attribute-based CRS \cite{CRS3,attribute1-crs,attribute2-crs} typically aims to enhance recommendation performance and reduce the number of dialogue turns required to complete recommendation tasks. They focus on asking clarifying questions \cite{zou2022learning,ma2024ask,zou2020question} and gradually identifying the best candidate set based on user preferences. For example, many studies typically employed reinforcement learning \cite{attribute1-crs,attribute2-crs,rl-1} or bandit-based approaches \cite{laohu}, to optimize the long-term benefits of asking clarifying questions.

Generation-based CRS \cite{hycorec, unicrs,DisenCRS, DCRS,zou2022seq1,zou2022seq2} emphasizes user interaction through natural language dialogue, intending to provide accurate recommendations and coherent responses, thereby enhancing its relevance to real-world application scenarios. For instance, \citet{KBRD} proposed a method that incorporates external knowledge to enhance recommendation and conversation effectiveness. \citet{KGSF} proposed entity-based and word-based knowledge graphs to enrich entity modeling and generate high-quality responses. \citet{c2crs} considered three types of data, i.e., reviews, knowledge graphs, and conversational contexts, and designed a coarse-to-fine contrastive learning approach to integrate these different data types. Besides knowledge graphs, \citet{unicrs} integrated recommendation and conversation tasks into an LLM through a unified prompt learning framework. \citet{DCRS} examined the application of semantically similar conversational contexts to enhance soft prompts in prompt learning. However, users' preferences are often based on their past multi-modal experiences, making the multi-modal features of items crucial for modeling user preferences. To this end, different from the aforementioned studies, we incorporate the multi-modal semantic information of items into CRS.

\begin{figure*}[htbp]
  \centering
  \includegraphics[width=\textwidth]{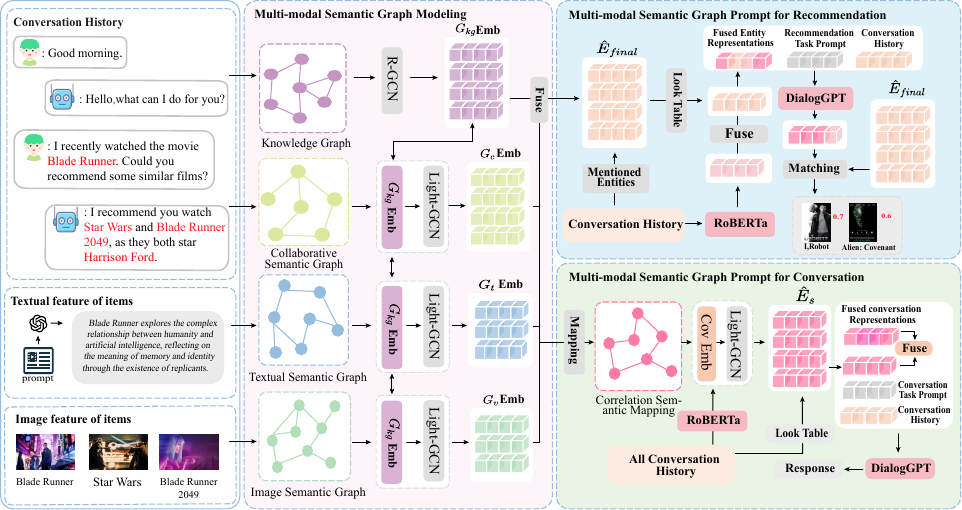}
  \caption{
  The framework of MSCRS.}
  \label{fig:MSCRS}
\end{figure*}
\subsection{Multi-modal Recommendation}
Multi-modal recommendation enhances performance by leveraging the multi-modal features of items. Early approaches \cite{VBPR,MMGCN} typically incorporate multi-modal features of items as a complement to ID features within the collaborative filtering framework. Due to the development of GNNs, an increasing number of studies \cite{MMGCN,MMSSL} combine multi-modal features with graphs. For example, \citet{MMGCN} proposed user and item representations in different modalities through specific modality graph structures. \citet{LATTICE} considered specific modality semantic graphs and integrated them with graph-based collaborative filtering methods. To address the semantic gap between different modalities, \citet{BM3} proposed a novel method and aligned these features through contrastive learning. Similarly, \citet{MMSSL} employed adversarial learning to learn user and item representations across different modalities and fuse these representations through cross-modal contrastive learning. Besides contrastive learning, \citet{promptmm} fused user preferences from different modalities through ranking distillation \cite{RD}. Unlike the approaches mentioned above, we generate various semantic graphs and combine them with prompt learning in the CRS scenario. This allows LLMs to understand the topological structure of GNNs, thereby guiding the generation of recommendation and conversation tasks for CRSs.

\section{PRELIMINARIES}
We denote the set of items by $\mathcal{I} = \{i_1, i_2, \ldots, i_N\}$ and the set of conversational contexts by $\mathcal{S} = \{s_1, s_2, \ldots, s_M\}$. In the conversations $\mathcal{S}$, we extract all entities involved into the set $\mathcal{E}=\{e_1, e_2, \ldots, e_K\}$, with $\mathcal{I} \subseteq \mathcal{E}$. $N$ is the number of items, $M$ is the number of conversational contexts, and $K$ is the number of all entities. Additionally, we collect the multi-modal features of items, denoted as $x_i^m \in \mathbb{R}^{d_m}$, where $d_m$ denotes the dimension of the feature, and $m\in\{t,v\}$, where $t$ denotes textual modality and $v$ denotes image modality.
A conversational context $s \in \mathcal{S}$ is represented as a collection of utterances $c$, expressed as $s = \{c_b\}_{b=1}^{n}$. In the \(b\)-th turn of the conversation, each utterance \(c_b\) consists of a sequence of words $w$, expressed as \(c_b = \{w_j\}_{j=1}^m\). The set of words is denoted by $\mathcal{W}$. As the conversation progresses, utterances are aggregated into a conversation history. CRS uses this history to infer user preferences and generate conversation responses. During the \( b \)-th turn, the recommender component recommends a set of candidate items from the complete item set \( \mathcal{I} \) based on the modeled user preferences. Meanwhile, the conversation component generates the next utterance \(c_{b}\) in response to the preceding conversation.

\section{METHODOLOGY}
Our approach consists of four main parts, as shown in Figure \ref{fig:MSCRS}. First, we extract the corresponding multi-modal data for items and then encode them. Second, we introduce the multi-modal semantic graph modeling component, which primarily integrates the proposed specific semantic graphs of multiple modalities. Finally, we elaborate on our methods for recommendation and conversation tasks through multi-modal semantic graph prompt learning.

\subsection{Multi-modal Feature Encoding}
As shown in Figure \ref{fig:MSCRS}, our model primarily considers three types of data: conversation history, textual descriptions of items, and image features of items. Next, we will introduce the feature extraction and encoding methods for each of these three types of data.

\noindent \textbf{Encoding Conversation History.}
Like previous work \cite{KGSF,c2crs}, we map the items (e.g., movies) and related entities (e.g., actors) in the conversation history to the knowledge graph DBpedia \cite{dbpedia} to capture the intricate interconnections between entities. By incorporating the knowledge graph DBpedia, we enhance the semantic information of these entities. The knowledge graph $\mathcal{G}_{kg}$ consists of a set of entities $\mathcal{E}$ and a set of edges $\mathcal{R}$. It uses triples $\langle e_1, r, e_2 \rangle$ to store semantic facts, where \(e_1, e_2 \in \mathcal{E}\) represent items or item-related entities, and \(r \in R\) denotes the relationship between \(e_1\) and \(e_2\). 
We apply R-GCN \cite{RGCN} for encoding \(\mathcal{G}_{kg}\). Specifically, the representation of node \( e \) at the \( (l + 1) \)-th layer is computed as:
\begin{equation}
\mathbf{n}_{e}^{(l+1)} = \sigma ( \sum_{r \in \mathcal{R}} \sum_{e' \in \mathcal{E}_e^r} \frac{1}{Z_{e,r}} \mathbf{W}_r^{(l)} \mathbf{n}_{e'}^{(l)} + \mathbf{W}^{(l)} \mathbf{n}_e^{(l)}),
\label{eq:1}
\end{equation}
where \( \mathbf{n}_e^{(l)} \) represents the embedding of node \( e \) at the \( l \)-th layer, and \( \mathcal{E}_e^r \) refers to the set of neighboring nodes for \( e \) associated with relation \( r \). The matrix \( \mathbf{W}^{(l)} \) applies a learnable transformation to the node embeddings at the \( l \)-th layer, and \( \mathbf{W}_r^{(l)} \) transforms the embeddings of neighboring nodes connected by relation \( r \) using a relation-specific matrix. The factor \( Z_{e,r} \) normalizes the contribution of each neighboring node. 
\begin{figure}[tbp]
  \centering
  \includegraphics[width=0.4\textwidth]{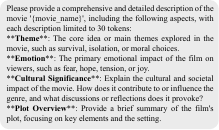}
  \caption{The prompt template.}
  \label{fig:temple}
\end{figure}

\noindent \textbf{Encoding Textual Descriptions of Items.} Based on the extensive general knowledge of existing LLMs, we employ the powerful GPT language model (GPT-4o) to extract textual descriptions $v_i^t$ for an item $i$:
\begin{equation}
v_i^t = \mathbf{F}_{\text{GPT}}(i; \theta_p),
\end{equation}
where $\theta_p$ is the prompt template, $v_i^t$ represents the text description generated by $\mathbf{F}_{\text{GPT}}$. The specific prompt template is provided in Figure \ref{fig:temple}. After generating the textual description $v_i^t$, we utilize the pre-trained model RoBERTa \cite{RoBERTa} to encode it:
\begin{equation}
x_i^t = \text{AvgPool}\left( \mathbf{F}_{\text{RoBERTa}}(v^t_i; \theta_r) \right),
\end{equation}
where AvgPool represents the average pooling operation, \( \theta_r \) denotes all the parameters of RoBERTa. Finally, we obtain \( X^t = \{ x_1^t, x_2^t, x_3^t, \ldots, x_N^t \} \), which denotes the textual features of all items.

\noindent \textbf{Encoding Image Features of Items.}
The visual features of items contain rich semantic information. We collect multiple still images \( \{m^v_1, m^v_2, \ldots, m^v_l\} \) of item \( i \) using web scraping from IMDb. Then, we extract the image representations corresponding to the item \( i \) using the pre-trained model ViT \cite{vit}:
\begin{equation}
Z_i = \{z^v_1, z^v_2, \ldots, z^v_l \} = \mathbf{F}_{\text{ViT}}(m^v_1, m^v_2, \ldots, m^v_l; \theta_v).
\end{equation}

Then, we calculate the average of these embeddings to obtain the image feature representation for each item \( i \):
\begin{equation}
x^v_i = \frac{1}{l} \sum_{j=1}^{l} z^v_j,
\end{equation}
where $l$ denotes the number of images of the item $i$, $x^v_i$ represents the image feature representation of item $i$. Finally, we obtain the image features of all items as \( X^v = \{ x_1^v, x_2^v, x_3^v, \ldots, x_N^v \} \).

After encoding, we can generate representations for the textual features of items, image features of items, and knowledge graph features of all entities. 
Next, we explore how to model and fuse these multi-modal features to obtain a unified representation.

\subsection{Multi-modal Semantic Graph Modelling}
\label{sec:3.2}
\noindent \textbf{Collaborative Semantic Graph.}
Although the knowledge graph models the complex real-world knowledge among global entities to some extent, it still faces issues such as noise, errors, inconsistent data, and ``data silos". These problems can impact the accuracy and reliability of downstream recommendation tasks. To address these challenges, we introduce the collaborative semantic graph. The collaborative semantic graph models the relationships between entities from a co-mention perspective, thereby enhancing the structural information of the entities. The collaborative semantic graph $\mathcal{G}_{c}=(\mathcal{E},\mathcal{R}_c)$, where $\mathcal{R}_c$ is the set of edges. The matrix $\textbf{C}\in \mathbb{R}^{K\times K}$ is a sparse matrix representing the co-mention counts between entities, where $K$ is the total number of entities. Specifically, the elements of the matrix are defined as:
\begin{equation}
\textbf{C}_{i,j} = \sum_{y=1}^{Y} \text{count}(e_i, e_j \mid Q_y),
\end{equation}
where $Q_y=[e_{\text{1}}, e_{2},\ldots, e_{K}](e \in \mathcal{E})$ denote the entities that appear in a single conversation, and $Y$ is the total number of conversational contexts in the train data. Each element \( \textbf{C}_{i,j} \) quantifies the frequency with which entities \( e_i \) and \( e_j \) are co-mentioned across all conversations, thereby revealing their potential associations. This matrix $\textbf{C}$ serves as the foundation for constructing the collaborative semantic graph \( \mathcal{G}_c \), where an edge is established between entities \( e_i \) and \( e_j \) if their co-mention count exceeds a predefined threshold, reflecting their semantic relationships.

As shown in Eq. (\ref{eq:1}), we adopt an embedding table \( \mathbf{N}^1 \in \mathbb{R}^{K \times d} \) generated by a layer of R-GCN as the initial embedding table \( \mathbf{E}^{(0)}_{c} \in \mathbb{R}^{K \times d} \) for the collaborative semantic graph \( \mathcal{G}_c \). Then, we utilize LightGCN \cite{lightgcn} for encoding \( \mathcal{G}_c \). LightGCN streamlines the graph convolution operations by omitting feature transformation and nonlinear activation components, enhancing recommendation effectiveness while also facilitating the optimization of the model. Specifically, the representations for items at the \( l \)-th layer of graph convolution in \( \mathcal{G}_c \) are derived as follows:
\begin{equation}
\label{eq7}
\mathbf{E}_c^{(l)} = (\mathbf{D}_c^{-\frac{1}{2}} \mathbf{C} \mathbf{D}_c^{-\frac{1}{2}}) \mathbf{E}_c^{(l-1)},
\end{equation}
where $\mathbf{D}_c \in \mathbb{R}^{K\times K}$ is the degree matrix. We obtain \( l \)-layer representations from the \( l \)-layer collaborative semantic graph, and then generate the average entity embedding table $\hat{E}_c$ using average pooling:
\begin{equation}
\hat{\mathbf{E}}_c= \text{AvgPool}([\mathbf{E}_c^{(0)}, \mathbf{E}_c^{(1)}, \ldots, \mathbf{E}_c^{(l)}]).
\end{equation}

\noindent \textbf{Textual and Image Semantic Graph.}
User experience with actual items often stems from multi-modal perception (e.g., when watching a movie, users not only focus on the plot and dialogue but are also influenced by visual effects and the soundtrack, which together shape their viewing experience). Meanwhile, the multi-modal features of items provide rich and valuable information. In this section, we propose modality-specific semantic graphs to comprehensively model the multi-modal features of items. Grounded in the idea that similar items are more inclined to interact than dissimilar items \cite{LATTICE}, we evaluate the semantic relationship between two items based on their similarity. In Section \ref{sec:3.2}, we obtain the text features \( \mathbf{X}^t \in \mathbb{R}^{N \times d_t} \) and image features \( \mathbf{X}^v \in \mathbb{R}^{N \times d_v} \) of the items. We calculate the semantic relevance between modality-specific features using cosine similarity:
\begin{equation}
\mathbf{A}_{i j}^{m}=\frac{(\boldsymbol{x}_{i}^{m})^{\top} \boldsymbol{x}_{j}^{m}}{\|\boldsymbol{x}_{i}^{m}\|\|\boldsymbol{x}_{j}^{m}\|},
\label{eq:9} 
\end{equation}
where \( m \in\{t, v\} \), $\mathbf{A}^{m}\in \mathbb{R}^{N \times N}$. A higher value of \( \mathbf{A}_{ij}^m \) indicates a stronger semantic correlation between items \( i \) and \( j \) within modality \( m \). Typically, the adjacency matrix of a graph is expected to be nonnegative; however, \( \mathbf{A}_{ij} \) spans the interval \([-1, 1]\). Consequently, we set the negative values to zero. Furthermore, common graph structures tend to be much sparser than fully connected graphs, which not only incurs higher computational costs but may also introduce extraneous and insignificant edges. We perform \(k\)NN sparsification \cite{knn} on the dense graph $A^{m}$. For each item \(i\), we retain only the top-\(k\) edges with the highest confidence scores:
\begin{equation}
\tilde{\mathbf{A}}^m_{ij}= 
\begin{cases} 
1 & \text{if } \mathbf{A}^m_{ij} \in \text{top-}k(\mathbf{A}^m_i), \\ 
0 & \text{otherwise},
\end{cases}
\label{eq:10} 
\end{equation}
where $\tilde{\mathbf{A}}^m_{ij}$ is the sparsified adjacency matrix. Similar to Eq. (\ref{eq7}), we adapt the LightGCN to encode the modality-specific semantic graph:
\begin{equation}
\mathbf{E}_m^{(l+1)} = (\mathbf{D}_m^{-\frac{1}{2}} \tilde{\mathbf{A}}^m \mathbf{D}_m^{-\frac{1}{2}}) \mathbf{E}_m^{(l)},
\label{eq:11} 
\end{equation}
where \( \mathbf{D}_m \in \mathbb{R}^{N \times N}\) denotes the degree matrix of the modality-specific semantic graph. Consistent with the method used to initialize $\mathcal{G}_c$, we initialize the modality-specific semantic graph embedding table $\mathbf{E}_m^{(0)} \in \mathbb{R}^{N \times d}$ using the embedding table \( \mathbf{N}^1 \in \mathbb{R}^{K \times d} \) enhanced by a layer of R-GCN. We obtain \( l \)-layer representations from the \( l \)-layer modality-specific semantic graph, and then generate the entity embedding table using average pooling:
\begin{align}
\hat{\mathbf{E}}_t &= \text{AvgPool}([\mathbf{E}_t^{(0)}, \mathbf{E}_t^{(1)}, \ldots, \mathbf{E}_t^{(l)}]), \\
\hat{\mathbf{E}}_v &= \text{AvgPool}([\mathbf{E}_v^{(0)}, \mathbf{E}_v^{(1)}, \ldots, \mathbf{E}_v^{(l)}]),
\end{align}
where $\hat{\mathbf{E}}_t$ is the average embedding table of textual semantic graph, while $\hat{\mathbf{E}}_v$ is the average embedding table of the image semantic graph. $\hat{\mathbf{E}}_t$ and $\hat{\mathbf{E}}_v$ capture semantic information from multiple layers of modality-specific semantic graphs, providing a more comprehensive representation of the items.

While the textual and image semantic graphs specifically enhance items, we fused the two modality-specific semantic graphs using a weighting function:
\begin{equation}
\hat{\mathbf{E}}_m = \lambda \hat{\mathbf{E}}_t + (1-\lambda) \hat{\mathbf{E}}_v,
\end{equation}
where $\lambda \in (0, 1)$ is the hyperparameter controling the fusion ratio.

Next, we fuse the original knowledge graph with the collaborative semantic graph:
\begin{equation}
\hat{\mathbf{E}}_\alpha =\text{AvgPool}([ \hat{\mathbf{E}}_c, \mathbf{N}^{(1)}]).
\end{equation}

Then, we fuse the multi-modal embeddings $\hat{E}_m$ and $\hat{\mathbf{E}}_\alpha$:
\begin{equation}
\hat{\mathbf{E}}_{final} = \hat{\mathbf{E}}_\alpha[\mathcal{I}] + \hat{\mathbf{E}}_m,
\end{equation}
where \(\mathcal{I}\) represents the indices of common items between \(\hat{\mathbf{E}}_\alpha\) and \(\hat{\mathbf{E}}_m\).

\subsection{Multi-modal Semantic Graph Prompt Learning For Recommendation}

In Section \ref{sec:3.2}, we generated the fused embeddings $\hat{\mathbf{E}}_{final}$ from $\hat{\mathbf{E}}_m$ and $\hat{\mathbf{E}}_\alpha$. 
For a conversation $s \in S$, we can query the embedding \( \mathbf{V}_s \in \mathbb{R}^{q \times d}\) of $q$ entities involved in the conversation \( s \) from \( \hat{\mathbf{E}}_{final} \). We use RoBERTa to extract the embedding \( \mathbf{\mathbf{T}}_s \in \mathbb{R}^{p \times d_c} \) of the current conversation \( s \), where \( p \) represents the number of tokens in the conversational context, and \( d_c \) represents the dimensionality of the token embeddings. Next, we map $\mathbf{V}_s$ to the same dimension as $\mathbf{T}_s$ using a bilinear function:
\begin{equation}
\tilde{\mathbf{V}}_s =\mathbf{W}_1 \mathbf{V}_s \mathbf{W}_2,
\end{equation}
where $\mathbf{W}_1 \in \mathbb{R}^{p \times q}$ and $\mathbf{W}_2 \in \mathbb{R}^{d \times d_c}$ are weight matrix.

We fuse entities $\tilde{\mathbf{V}}_s$ and conversation $\mathbf{T}_s$ using a contrastive learning: 
\begin{align} 
L_{fuse}=-(log\frac{exp(\mathbf{T}_s\cdot {\tilde{\mathbf{V}}_s}/\tau )}{\sum_{\gamma}^{\Omega }exp(\mathbf{T}_s\cdot {\mathbf{T}_\gamma}/\tau ) } +log\frac{exp(\tilde{\mathbf{V}}_s\cdot {\mathbf{T}_s}/\tau )}{\sum_{\gamma }^{\Omega }exp(\tilde{\mathbf{V}}_s\cdot {\tilde{\mathbf{V}}_\gamma }/\tau )}),
\end{align}
where $\Omega$ denotes the number of negative examples of contrastive learning and $\tau$ denotes the temperature coefficient. The final fused entity ${\hat{\mathbf{V}}_s}=\tilde{\mathbf{V}}_s+ \mathbf{T_s}$.

We adopt prompt learning \cite{prefix} to make use of LLM in a simple and flexible way. The final prompt $\mathbf{r}_s$ for the recommendation task consists of the following three parts:
\begin{equation}
\mathbf{r}_s=[\hat{ \mathbf{V}}_s; \mathbf{O}_{rec}; s],
\end{equation}
where $\mathbf{O}_{rec}$ denotes embeddings for the recommendation task-specific soft prompt (random initialization), and $s$ denotes the conversational context (word tokens). We chose DialoGPT \cite{dialogpt} as the base LLM, which uses a Transformer-based architecture and was pre-trained on a large-scale conversation corpus extracted from Reddit, as done in existing studies \cite{unicrs,DCRS}. 
We input \( \mathbf{r}_s \) into DialoGPT and apply a pooling layer to derive the multi-modal semantic graph enhanced user preference embedding \( \hat{\mathbf{r}}_s \in \mathbb{R}^d \):
\begin{equation}
\hat{\mathbf{r}}_s  = \text{Pooling}( \mathbf{F}_{\text{DialoGPT}}(\mathbf{r}_s; \theta_{rec}) ),
\end{equation}
where $\theta_{rec}$ denotes the trainable parameters, which consist of $\hat{ \mathbf{V}}_s$ and $\mathbf{O}_{rec}$. We use the last token representation of DialoGPT for item recommendations or generation tasks. 

\noindent \textbf{Pre-training.} Due to the semantic gap between the multi-modal semantic graph enhanced prompt $\hat{\mathbf{V}}_s$ and the conversational context $s$, we associate them through pre-training. Specifically, we employ multi-modal semantic graph enhanced user preference embedding \( \hat{\mathbf{r}}_s \) to predict the entities contained in the current conversation $s$. The probability of entity $i$ is calculated as follows:
\begin{equation}
\label{eq23}
\textbf{P}_{entity}(i) = \text{Softmax}(\hat{\mathbf{r}}_s \hat{\textbf{E}}^\top_{final}).
\end{equation}  

We combine the fuse loss with cross-entropy to optimize the model parameters.
\begin{equation}
L_{pre}(\theta_{rec})=-\sum_{j \in S_{train}} \sum_{i=1}^{K} \log \textbf{P}_{entity}^{j}(i \mid \hat{\mathbf{r}}_s, \theta_{rec}) + \delta L_{fuse},
\end{equation}
where $K$ is the number of entities, $S_{train}$ denotes the set of all conversations in the training set. $\delta$ is the hyper-parameter to control the fuse loss weight.

\noindent \textbf{Recommendation.} The recommendation task predicts the probabilities of all items. Similar to Eq. (\ref{eq23}), we generate the probabilities $\textbf{P}_{item}(i)$ of the recommended items:

\begin{equation}
\textbf{P}_{item}(i) = \text{Softmax}(\hat{\mathbf{r}}_s \hat{\textbf{E}}^\top_{final}[\mathcal{I}]).
\end{equation}  

Then we train the recommendation task using a cross-entropy loss and fuse loss:
\begin{equation}
L_{rec}(\theta_{rec})=-\sum_{j \in S_{train}} \sum_{i=1}^{N} y^j_{i}\log \textbf{P}_{item}^{j}(i \mid \hat{\mathbf{r}}_s, \theta_{rec}) + \delta L_{fuse},
\end{equation}
where $y_{i}^{j}$ is the corresponding label of the item \( i \) in the conversation instance \( j \), and $N$ is the number of items.

\subsection{Multi-modal Semantic Graph Prompt Learning For Conversation}
The conversation task aims to provide appropriate responses based on the current user utterance. Previously, we modeled the multiple relationships between different entities. By leveraging the entities present in various conversational contexts, we designed a correlation semantic mapping that integrates the contexts of all conversations in the training data. This approach allows us to capture conversational contexts with semantic similarities to the current conversational context, thereby enhancing the semantic information of the current conversational context.

Specifically, for a conversational context \( s \), let \( Q_s \) be the set of entities involved. We obtain the multi-modal semantic graph enhanced entity set $\hat{Q}_s$ through the first-order adjacency relationships of four types of semantic graphs (knowledge graph $\mathcal{G}_{kg}$, collaborative semantic graph $\mathcal{G}_c$, textual semantic graph $\mathcal{G}_t$, and image semantic graph $\mathcal{G}_v$). If there exists an edge connection between entities, we will also include the entities connected by these edges into the enhanced entity set $\hat{Q}_s$: 
\begin{equation}
\hat{Q}_s = Q_s \cup \bigcup_{G \in \{\mathcal{G}_{kg}, \mathcal{G}_c, \mathcal{G}_t, \mathcal{G}_v\}} \{ e_j \mid \exists e_i \in E(G), A_{ij}(G) = 1 \}.
\end{equation}

After obtaining the multi-modal semantic graph enhanced entity set $\hat{Q}$ for all conversational contexts $S$, we represent the similarity between different conversational contexts by the number of common entities. Similar to Eqs. (\ref{eq:9}, \ref{eq:10}, \ref{eq:11}), we construct the correlation semantic mapping \( \mathcal{G}_{s} \). 
The final enhanced representation of the conversational context based on the correlation semantic mapping is as follows:
\begin{equation}
\hat{\textbf{E}}_s= \text{AvgPool}([\textbf{E}_s^{(0)}, \textbf{E}_s^{(1)}]),
\end{equation}
where \( \textbf{E}_s^{(0)} \in \mathbb{R}^{(p \times d_c)\times M}  \) is initialized by encoding all conversational contexts into embeddings with RoBERTa, where $M$ is the number of all conversational contexts. We utilize MLP to simulate the neighbor aggregation of one layer of LightGCN and generate the enhanced representations $\hat{\textbf{E}}_s$. For a conversational context \( s \in S \), we fuse \( \hat{ \mathbf{T}}_s = \mathbf{T} +\tilde{\mathbf{V}}_s \) and \( \hat{e}_s \in \hat{\textbf{E}}_s \) to generate the enhanced context embedding $\tilde{\mathbf{T}}_s$:
\begin{equation}
\tilde{\mathbf{T}}_s = \text{AvgPool}([\text{Reshape}(\hat{e}_s),\hat{ \mathbf{T}}_s]).
\end{equation}
The final prompt $c_s$ for the conversation task consists of the following three parts:
\begin{equation}
\mathbf{c}_s=[\tilde{ \mathbf{T}}_s; \mathbf{O}_{cov}; s],
\end{equation}
where $\mathbf{O}_{cov}$ denotes the conversation task specific soft prompt embeddings (random initialization), $s$ denotes the conversational context (word tokens). Then we input \( \mathbf{c}_s \) into DialoGPT and apply a pooling layer to derive embedding \( \hat{\mathbf{c}}_s \). We use \( \hat{\mathbf{c}}_s \) to drive the loss for learn $\theta_{gen}$. The optimization function for the conversation task is shown as follows:
\begin{equation}
L_{g e n}(\theta_{g e n}) =-\frac{1}{Y} \sum_{i \in S_{train}} \sum_{j=1}^{l_{i}} \log \textbf{P}(w_{i, j} \mid \hat{\mathbf{c}}_s ; \theta_{gen} ; w_{<j})
\end{equation}
where $Y$ is the number of training contexts, \( l_i \) represents the length of the label utterance, and \( w_{< j} \) denotes the words preceding the \( j \)-th position.

\section{EXPERIMENTS}
\subsection{Experimental Setup}
\subsubsection{Datasets.} We validate the effectiveness of our model on two widely used conversation recommendation datasets, \textbf{ReDial} \cite{CRS2} and \textbf{INSPIRED} \cite{inspired}, similar to previous work \cite{unicrs, DCRS}. Both datasets are specifically designed for conversational movie recommendation, consisting of realistic conversations between users and agents about movie recommendations. We obtain movie stills from IMDb. 
Detailed statistics are presented in Table \ref{tab:dataset}.


\begin{table}[t]
\caption{Statistics of the used datasets in our experiments.}
\label{tab:dataset}
\centering
\small
\begin{tabular}{lccc}
\toprule
\textbf{Dataset} & \textbf{\# Conversations} & \textbf{\# Utterances} & \textbf{\# Entities/Items}\\
\midrule
ReDial & 10,006 & 182,150 & 64,364/6,924\\
INSPIRED & 1,001 & 35,811 & 17,321/1,123\\
\bottomrule
\end{tabular}
\end{table}
\begin{table*}[htbp]
\caption{Recommendation performance comparison on ReDial and INSPIRED datasets, with the best results in bold and * indicating significant improvements over the best baseline ($p$-value $<0.05$). Unless otherwise stated, * marks significant improvements and bold values denote the best performances in the following paper.}
\label{tab:rec}
\centering
\resizebox{\textwidth}{!}{
\begin{tabular}{c|ccc|cc|cc|ccc|cc|cc}
\toprule
 & \multicolumn{7}{c|}{\textbf{ReDial}} & \multicolumn{7}{c}{\textbf{INSPIRED}} \\
\cmidrule(lr){2-8} \cmidrule(lr){9-15}
\multirow{1}{*}{\textbf{Model}} & \multicolumn{3}{c|}{\textbf{Recall}} & \multicolumn{2}{c|}{\textbf{NDCG}} & \multicolumn{2}{c|}{\textbf{MRR}} & \multicolumn{3}{c|}{\textbf{Recall}} & \multicolumn{2}{c|}{\textbf{NDCG}} & \multicolumn{2}{c}{\textbf{MRR}} \\
\cmidrule(lr){2-4} \cmidrule(lr){5-6} \cmidrule(lr){7-8} \cmidrule(lr){9-11} \cmidrule(lr){12-13} \cmidrule(lr){14-15}
 & \textbf{@1} & \textbf{@10} & \textbf{@50} & \textbf{@10} & \textbf{@50} & \textbf{@10} & \textbf{@50} & \textbf{@1} & \textbf{@10} & \textbf{@50} & \textbf{@10} & \textbf{@50} & \textbf{@10} & \textbf{@50} \\
\midrule
Popularity     & 0.011 & 0.053 & 0.183 & 0.029 & 0.057 & 0.021 & 0.027 & 0.031 & 0.155 & 0.322 & 0.085 & 0.122 & 0.064 & 0.071 \\
TextCNN   & 0.010 & 0.066 & 0.187 & 0.033 & 0.059 & 0.023 & 0.028 & 0.025 & 0.119 & 0.245 & 0.066 & 0.094 & 0.050 & 0.056 \\
BERT       & 0.027 & 0.142 & 0.307 & 0.075 & 0.112 & 0.055 & 0.063 & 0.049 & 0.189 & 0.322 & 0.112 & 0.141 & 0.088 & 0.095 \\
\midrule
ReDial    & 0.010 & 0.065 & 0.182 & 0.034 & 0.059 & 0.024 & 0.029 & 0.009 & 0.048 & 0.213 & 0.023 & 0.059 & 0.015 & 0.023 \\
KBRD     & 0.033 & 0.150 & 0.311 & 0.083 & 0.118 & 0.062 & 0.070 & 0.042 & 0.135 & 0.236 & 0.088 & 0.109 & 0.073 & 0.077 \\
KGSF      & 0.035 & 0.175 & 0.367 & 0.094 & 0.137 & 0.070 & 0.079 & 0.051 & 0.132 & 0.239 & 0.092 & 0.114 & 0.079 & 0.083 \\
TREA    & 0.045 & 0.204 & 0.403 & 0.114 & 0.158 & 0.087 & 0.096 & 0.047 & 0.146 & 0.347 & 0.095 & 0.132 & 0.076 & 0.087 \\
COLA   & 0.048 & 0.221 & 0.426 & - & - & 0.086 & 0.096 & - & - & - & - & - & - \\
VRICR    & 0.054 & 0.244 & 0.406 & 0.138 & 0.174 & 0.106 & 0.114 & 0.043 & 0.141 & 0.336 & 0.091 & 0.134 & 0.075 & 0.085 \\
UNICRS   & 0.065 & 0.241 & 0.423 & 0.143 & 0.183 & 0.113 & 0.125 & 0.085 & 0.230 & 0.398 & 0.149 & 0.187 & 0.125 & 0.133 \\
DCRS & 0.076& 0.253 & 0.439 & 0.154 & 0.195 & 0.123 & 0.132 & 0.093 & 0.226 & 0.414 & 0.153 & 0.192 & 0.130 & 0.137 \\
\midrule
\textbf{MSCRS}  & \textbf{0.081*} & \textbf{0.264*} & \textbf{0.451*} & \textbf{0.161*} & \textbf{0.201*} & \textbf{0.128*} & \textbf{0.136*} & \textbf{0.096*}& \textbf{0.257*} & \textbf{0.425*} & \textbf{0.168*} & \textbf{0.202*} & \textbf{0.140*} & \textbf{0.148*} \\
\bottomrule
\end{tabular}
}
\end{table*}
\subsubsection{Baselines.} The CRS includes two tasks: recommendation and conversation. Consequently, we compare our method with the following representative methods:
\begin{itemize}[leftmargin=*]
\item \underline{\textbf{Popularity}}: A simple metric that ranks items based on their occurrence frequency in the dataset.
\item \underline{\textbf{BERT}} \cite{bert}: An extensively utilized pre-trained model designed for text classification applications. We fine-tune BERT to forecast a selection of potential items.
\item \underline{\textbf{DialogGPT}} \cite{dialogpt}: It is a large-scale generative pre-trained model trained on extensive dialogue data, specifically optimized for generating contextually relevant and fluent conversational responses
\item \underline{\textbf{GPT-2}} \cite{gpt2}: A powerful benchmark for text generation that benefits from extensive pre-training on language models.
\item \underline{\textbf{BART}} \cite{bart}: A denoising autoencoder pretraining model for generation tasks.
\item \underline{\textbf{Redial}} \cite{CRS2}: This model was introduced alongside the ReDial dataset, which includes an autoencoder for recommendations and a generation model based on hierarchical RNN.
\item \underline{\textbf{KBRD}} \cite{KBRD}: The method enhances recommendation and conversation tasks by introducing an entity-based knowledge graph.
\item \underline{\textbf{KGSF}} \cite{KGSF}: This method enhances the information of entities and words through entity-based and word-based knowledge graphs.
\item \underline{\textbf{TREA}} \cite{TREA}: This method models recommendation and conversation tasks through a multi-layer inferable tree structure.
\item \underline{\textbf{COLA}} \cite{COLA}: It enhances conversational recommendation systems by enriching item and user representations through an interactive user-item graph and retrieving similar conversations.
\item \underline{\textbf{VRICR}} \cite{VRICR}: This method enhances the original knowledge graph through dynamic inference using variational Bayes.
\item \underline{\textbf{UNICRS}} \cite{unicrs}: This method combines the recommendation and conversation sub-tasks into the same prompt learning paradigm.
\item \underline{\textbf{DCRS}} \cite{DCRS}: This method enhances recommendation and conversation tasks by retrieving conversations that are similar to the current conversation.
\end{itemize}

\subsubsection{Evaluation Metrics} 
We evaluate the recommendation and conversation tasks using different metrics. For recommendation, we follow \cite{DCRS,unicrs} and adopt \textbf{Recall@k} (k=1, 10, 50), \textbf{NDCG@k} (k=10, 50), and \textbf{MRR@k} (k=10, 50). For conversation, we apply both automatic and human evaluations. Automatic metrics include \textbf{BLEU-N} (N=2, 3), \textbf{ROUGE-N} (N=2, L) and \textbf{Distinct-N} (N=2, 3, 4). For human evaluation, we randomly select 100 responses from each model and ask three annotators to score them on \textbf{Fluency} and \textbf{Informativeness} (0–2 scale), then we compute the average score for all test samples.
\begin{table*}[tbp]
\centering
\setlength{\tabcolsep}{4pt} 
\caption{Automatic evaluation for the conversation task on Redial and INSPIRED datasets.}
\label{tab:cov}
\begin{tabular}{c|cc|cc|ccc|cc|cc|ccc}
\toprule
 & \multicolumn{7}{c|}{\textbf{ReDial}} & \multicolumn{7}{c}{\textbf{INSPIRED}} \\
\cmidrule(lr){2-8} \cmidrule(lr){9-15}
\multirow{1}{*}{\textbf{Model}} & \multicolumn{2}{c|}{\textbf{BLEU}} & \multicolumn{2}{c|}{\textbf{ROUGE}} & \multicolumn{3}{c|}{\textbf{DIST}} & \multicolumn{2}{c|}{\textbf{BLEU}} & \multicolumn{2}{c|}{\textbf{ROUGE}} & \multicolumn{3}{c}{\textbf{DIST}} \\

\cmidrule(lr){2-3} \cmidrule(lr){4-5} \cmidrule(lr){6-8} \cmidrule(lr){9-10} \cmidrule(lr){11-12} \cmidrule(lr){13-15}
 & \textbf{-2} & \textbf{-3} & \textbf{-2} & \textbf{-L} & \textbf{-2} & \textbf{-3} & \textbf{-4} &\textbf{-2} & \textbf{-3} & \textbf{-2} & \textbf{-L} & \textbf{-2} & \textbf{-3}& \textbf{-4}  \\
\midrule
    DialogGPT & 0.041 & 0.021 & 0.054 & 0.258 & 0.436 & 0.632 & 0.771 & 0.031 & 0.014 & 0.041 & 0.207 & 1.954 & 2.750 & 3.235 \\
    GPT-2& 0.031 & 0.013 & 0.041 & 0.244 & 0.405 & 0.603 & 0.757 & 0.026 & 0.011 & 0.034 & 0.212 & 2.119 & 3.084 & 3.643 \\
    BART & 0.024 & 0.011 & 0.031 & 0.229 & 0.432 & 0.615 & 0.705 & 0.018 & 0.008 & 0.025 & 0.208 & 1.920 & 2.501 & 2.670 \\
    \midrule
    ReDial & 0.004 & 0.001 & 0.021 & 0.187 & 0.058 & 0.204 & 0.442 & 0.001 & 0.000 & 0.004 & 0.168 & 0.359 & 1.043 & 1.760 \\
    KBRD & 0.038 & 0.018 & 0.047 & 0.237 & 0.070 & 0.288 & 0.488 & 0.021 & 0.007 & 0.029 & 0.218 & 0.416 & 1.375 & 2.320 \\
    KGSF& 0.030 & 0.012 & 0.039 & 0.244 & 0.061 & 0.278 & 0.515 & 0.023 & 0.007 & 0.031 & 0.228 & 0.418 & 1.496 & 2.790\\
    VRICR& 0.021 & 0.008 & 0.027 & 0.137 & 0.107 & 0.286 & 0.471 & 0.011 & 0.001 & 0.025 & 0.187 & 0.853 & 1.801 & 2.827\\
    TREA& 0.022 & 0.008 & 0.039 & 0.175 & 0.242 & 0.615 & 1.176 & 0.013 & 0.002 & 0.027 & 0.195 & 0.958 & 2.565 & 3.411\\
    COLA& 0.026 & 0.012 & - & - & 0.387 & 0.528 & 0.625 & - & - & - & - & - & -\\
    UNICRS& 0.045 & 0.021 & 0.058 & 0.285 & 0.433 & 0.748 & 1.003 & 0.022 & 0.009 & 0.029 & 0.212 & 2.686 & 4.343 & 5.520 \\
    DCRS& 0.048 & 0.024 & 0.063 & 0.285 & 0.779 & 1.173 & 1.386 & 0.033 & 0.014 & 0.045 & 0.229 & 3.950 & 5.729 & 6.233  \\
    \midrule
   \textbf{MSCRS} & \textbf{0.054*} & \textbf{0.027*} & \textbf{0.070*} & \textbf{0.294*} 
   & \textbf{0.784*} & \textbf{1.332*} & \textbf{1.553*}
   &\textbf{0.040*} & \textbf{0.019*}  & \textbf{0.052*} & \textbf{0.235*} & \textbf{4.197*} & \textbf{5.983*} & \textbf{6.556*}\\    
    \bottomrule
\end{tabular}
\end{table*}

\subsubsection{Implementation Details}
We trained our proposed model on a 32GB V100 GPU. In our model, we use RoBERTa \cite{RoBERTa} to encode the textual features of items and input tokens. We employ ViT \cite{vit} to extract the image features of movies, and use DialoGPT-small~\cite{dialogpt} as the base LLM. The extracted textual features have a dimensionality of 768, while the image features have a dimensionality of 1024. We then map the features of different modalities to the same dimension. We set the number of R-GCN \cite{RGCN} layers to 1. For the collaborative semantic graph, text semantic graph, and image semantic graph, we use a 3-layer of LightGCN \cite{lightgcn} for encoding. We tune our soft prompt between 10 and 50 for the recommendation and conversation tasks. The batch size is set to 64 for the recommendation task and 8 for the conversation task. We use Adam \cite{adam} as the optimizer for our model and adjust our learning rate between 1e-5 and 1e-3.

\subsection{Evaluation on Recommendation Task}

\subsubsection{Automatic Evaluation.} Table \ref{tab:rec} shows the experimental results for the recommendation task. Our MSCRS model achieves state-of-the-art performance, ranking first across all metrics on ReDial and INSPIRED. Compared to the strongest baseline DCRS, MSCRS achieves 0.081 (+6.5\% for ReDial) and 0.096 (+3.2\% for INSPIRED) in Recall@1, indicating higher accuracy in satisfying users immediate needs. For Recall@10, MSCRS achieves 0.264 (+4.3\% for ReDial) and 0.257 (+13.7\% for INSPIRED), demonstrating superior top-10 coverage. Additionally, Recall@50 scores of 0.451 (+2.7\% for ReDial) and 0.425 (+2.6\% for INSPIRED) show that MSCRS maintains high accuracy in longer recommendation lists.

Compared to knowledge-enhanced CRS models (KBRD, KGSF, COLA, VRICR), MSCRS significantly improves recommendation accuracy by leveraging multi-modal semantic relationships. Against LLM-based CRS models (UNICRS, DCRS), MSCRS outperforms UNICRS and DCRS on all metrics. The superior performance of the MSCRS model can be attributed to its ability to learn rich knowledge from three different semantic structures: the collaborative semantic graph, the textual semantic graph, and the image semantic graph. By combining these graph structures with higher-order semantic relationships and prompt learning for LLMs, the MSCRS model achieves significantly better results.

\subsubsection{Ablation Study.} Our recommendation method mainly enhances embeddings of items and item-related entities through the collaborative semantic graph, text semantic graph, and image semantic graph. To explore their impact on model performance, we designed four ablation variants: (1) \textbf{MSCRS w/o -co}, removing the collaborative semantic graph; (2) \textbf{MSCRS w/o -t}, removing the text semantic graph; (3) \textbf{MSCRS w/o -i}, removing the image semantic graph; and (4) \textbf{MSCRS w/o -r}, removing all three graph structures. Recall@10 is used as the evaluation metric due to its simplicity and consistent trends with Recall@1 and Recall@50.

From Figure \ref{fig:combined} (a) and (b), it can be seen that removing any of the three graphs results in drops in performance. Removing the collaborative semantic graph (MSCRS w/o -co) resulted in a decrease in Recall@10, indicating its crucial role in capturing relationships between entities. Removing the textual semantic graph (MSCRS w/o -t) and the image semantic graph (MSCRS w/o -i) also results in a decline in performance, emphasizing the importance of textual and image information for recommendation quality. Finally, the variant that removes all enhanced graph structures (MSCRS w/o -r) exhibits the lowest Recall@10 value, further demonstrating the necessity of combining multiple graph structures to improve model performance. In summary, the ablation study shows that all the collaborative semantic graph, textual semantic graph, and image semantic graph improve the effectiveness of the recommendation.

\begin{table}[tbp]
\centering
\caption{Human evaluation for the conversation task on Redial dataset.}
\label{tab:performance}
\begin{tabular}{lcc}
    \toprule
    \textbf{Models} & \textbf{Fluency} & \textbf{Informativeness} \\
    \midrule
    ReDial & 1.31 & 0.98 \\
    KGSF & 1.21 & 1.16 \\
    \midrule
    GPT-2 & 1.56 & 1.52\\
    BART & 1.48 & 1.43 \\
    UNICRS & 1.68 & 1.56 \\
    DCRS & 1.74 & 1.62 \\ 
    \midrule
    \textbf{MSCRS} & \textbf{1.79*} & \textbf{1.67*} \\
    \bottomrule
\end{tabular}
\end{table}

\subsection{Evaluation on Conversation Task}
\subsubsection{Automatic Evaluation.} Table \ref{tab:cov} presents the comparison of BLEU, ROUGE, and DIST scores on the ReDial and INSPIRED datasets. On the all dataset, MSCRS achieves the best performance across all three metrics, showcasing its superior text generation capability. These results highlight the outstanding conversation generation performance of MSCRS across both datasets, which can be attributed to its ability to establish complex high-order associations between enhanced entity representations and text during pre-training. Furthermore, the proposed correlation semantic mapping effectively enriches the semantic context, enabling MSCRS to generate more informative and fluent responses.

\subsubsection{Human Evaluation.} Table \ref{tab:performance} indicates that MSCRS excels in fluency and informativeness, showcasing its robust capability to generate high-quality conversations. This is likely due to our multi-modal semantic awareness, which enhances conversation generation quality through more complex relationship modeling. ReDial achieves the lowest scores in both metrics. KGSF shows an improvement in informativeness but performs slightly worse in fluency, suggesting progress in content richness but a need for further improvement in language naturalness. GPT-2 and BART perform well in fluency and informativeness, validating the effectiveness of pre-training techniques in natural language generation tasks. UNICRS and DCRS outperform GPT-2 and BART in both metrics, with both providing optimization approaches for the conversation generation task, thereby demonstrating their capabilities in generating high-quality conversations.

\begin{figure}[tbp]
    \centering
    \begin{minipage}{0.23\textwidth}
        \centering
        \includegraphics[width=\textwidth]{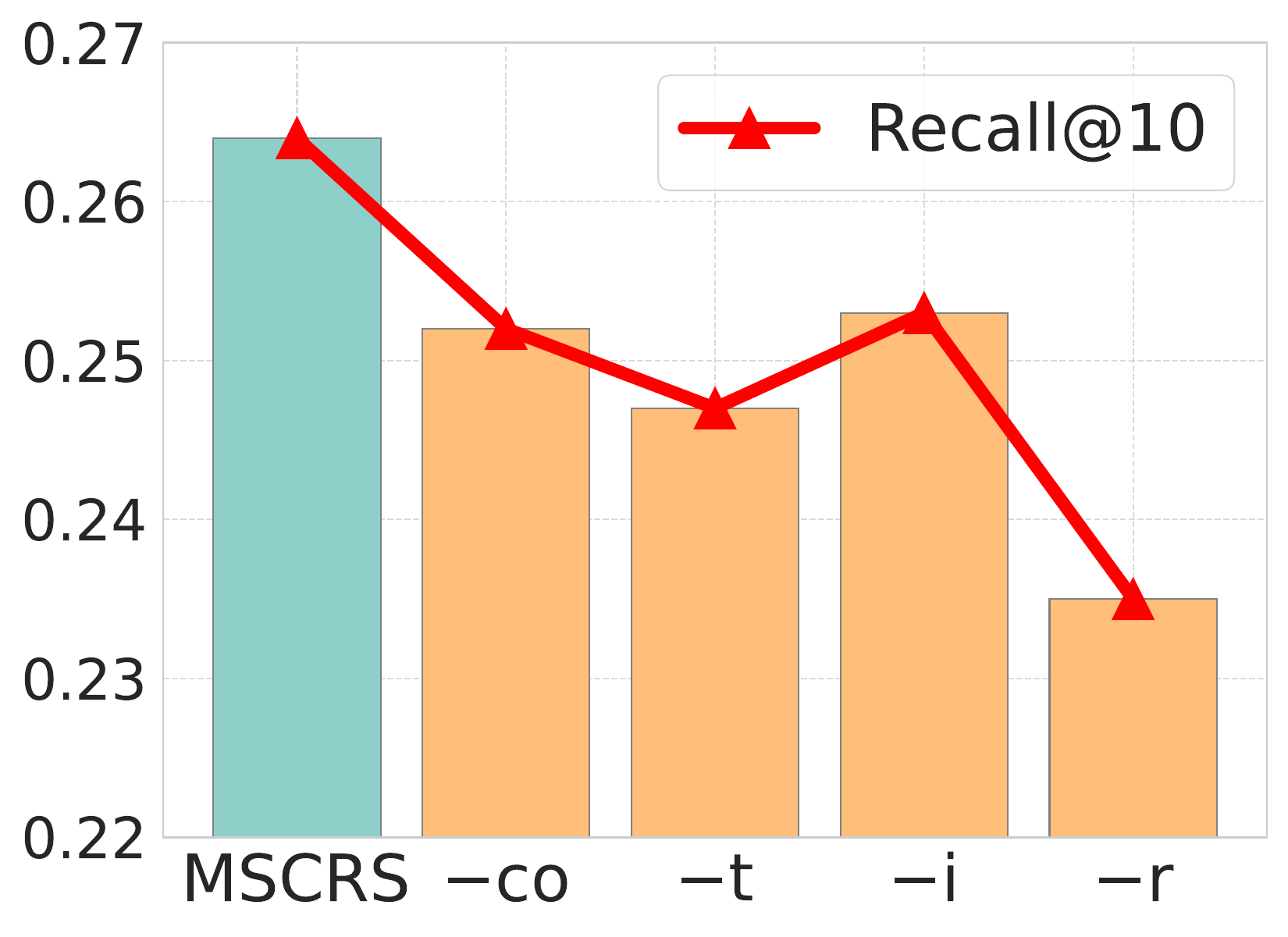}
        \caption*{(a) ReDial (Rec.)}
    \end{minipage}\hfill
    \begin{minipage}{0.23\textwidth}
        \centering
        \includegraphics[width=\textwidth]{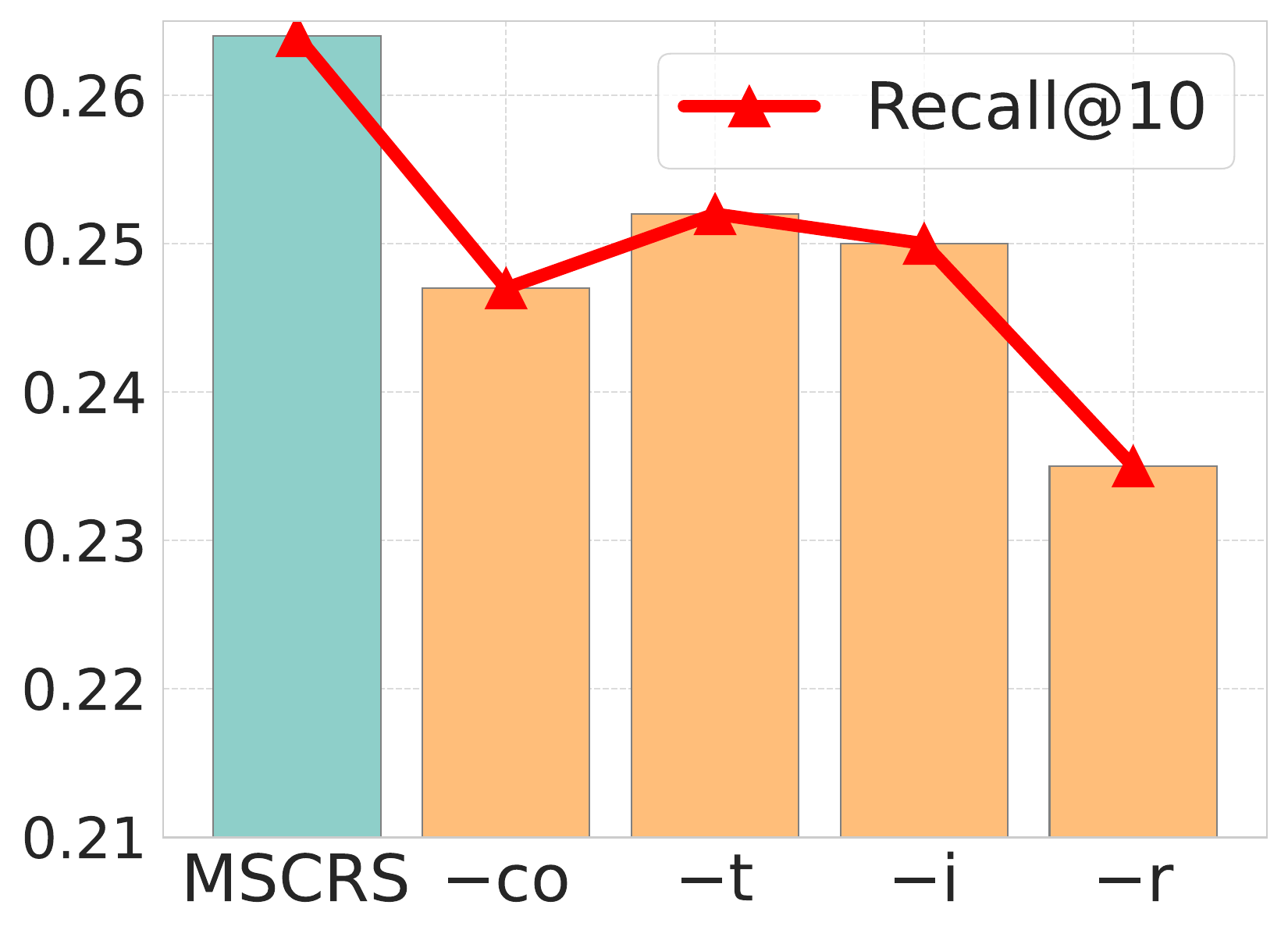}
        \caption*{(b) INSPIRED (Rec.)}
    \end{minipage}
    
    \begin{minipage}{0.23\textwidth}
        \centering
        \includegraphics[width=\textwidth]{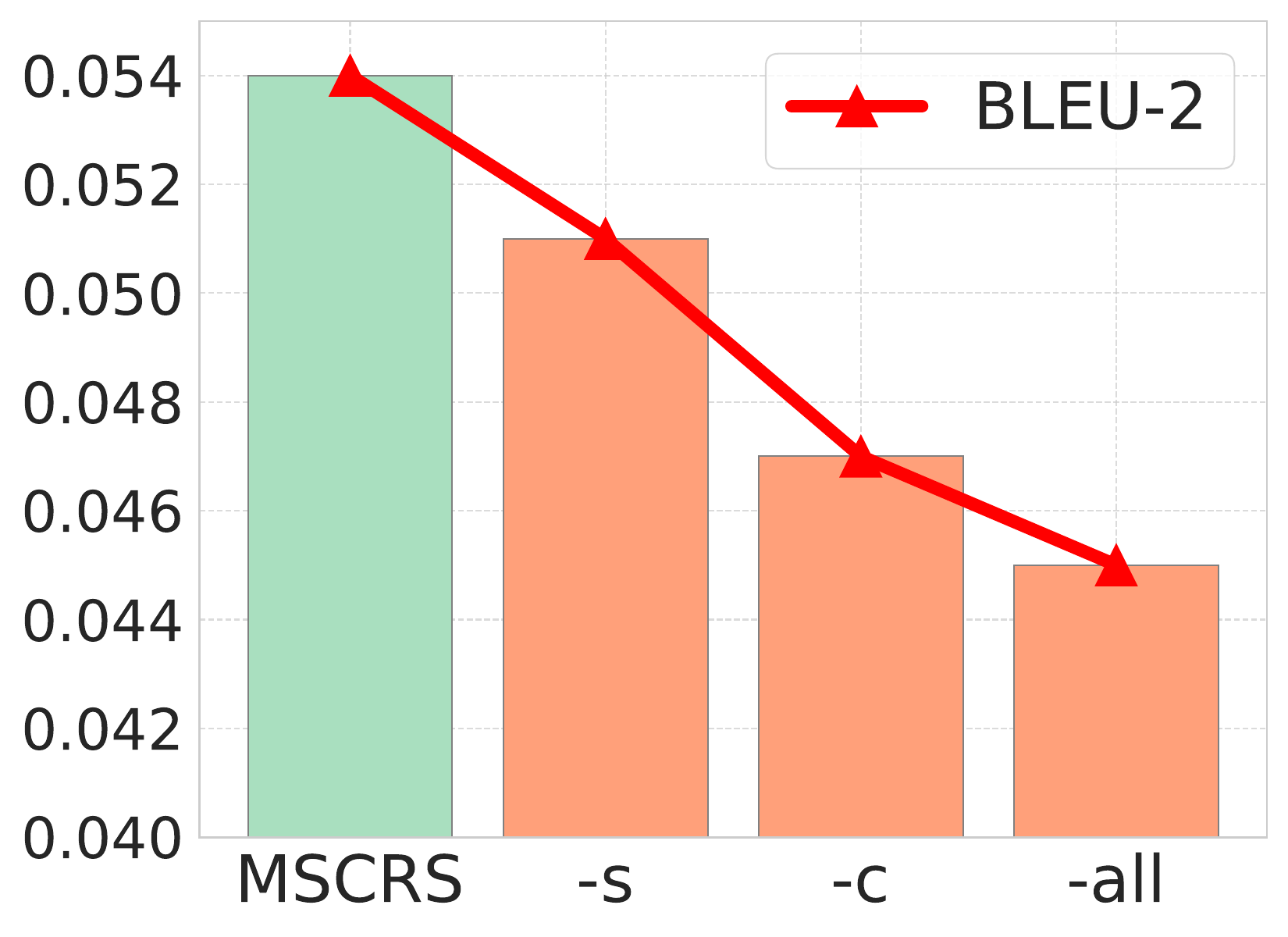}
        \caption*{(c) ReDial (Cov.)}
    \end{minipage}\hfill
    \begin{minipage}{0.23\textwidth}
        \centering
        \includegraphics[width=\textwidth]{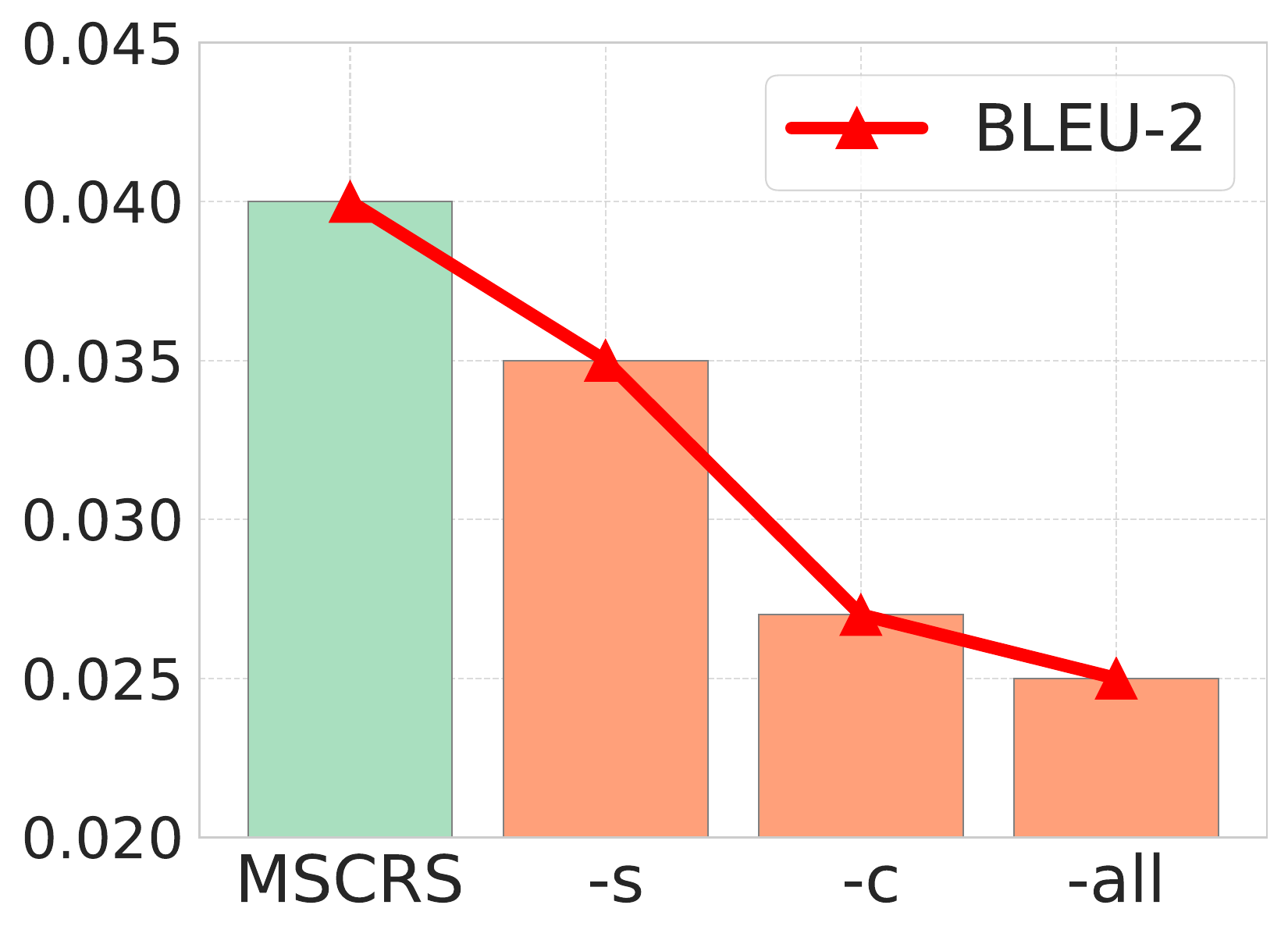}
        \caption*{(d) INSPIRED (Cov.)}
    \end{minipage}
    
    \caption{Ablation studies for the recommendation and conversation tasks on the ReDial and INSPIRED datasets.}
    \label{fig:combined}
\end{figure}

\subsubsection{Ablation Study.}
Our proposed model enhances response generation primarily through multi-modal semantic graph enhanced entity and correlation semantic mapping. To verify the effectiveness of these two modules, we designed different variants for ablation experiments: (1) \textbf{MSCRS w/o -s} indicates that removing our proposed multi-modal semantic enhanced entity; (2) \textbf{MSCRS w/o -c} indicates removing the correlation semantic mapping enhancement component; (3) \textbf{MSCRS w/o -all} indicates the removal of both the multi-modal semantic graph enhanced entity component and the correlation semantic mapping component. 

From Figure \ref{fig:combined} (c) and (d), we observe that in both datasets, MSCRS w/o -s leads to a decline in the BLEU-2 score, indicating that multi-modal semantic graph enhanced entities improve the quality of response generation. Similarly, MSCRS w/o -c also results in a more significant performance decline, demonstrating the importance of correlation semantic mapping in enhancing conversational context and generating coherent and fluent conversation. The worst performance is observed with MSCRS w/o -all, further validating the indispensable role of these two modules in jointly enhancing the model in conversation generation. These experiments verify the independent effectiveness of the multi-modal semantic graph enhanced entity components and the correlation semantic mapping components.
 
\subsection{Further Analysis.}
\subsubsection{Effect of $k$}
\begin{figure}[tbp]
    \centering
    \begin{minipage}[t]{0.23\textwidth}
        \centering
        \includegraphics[width=\textwidth]{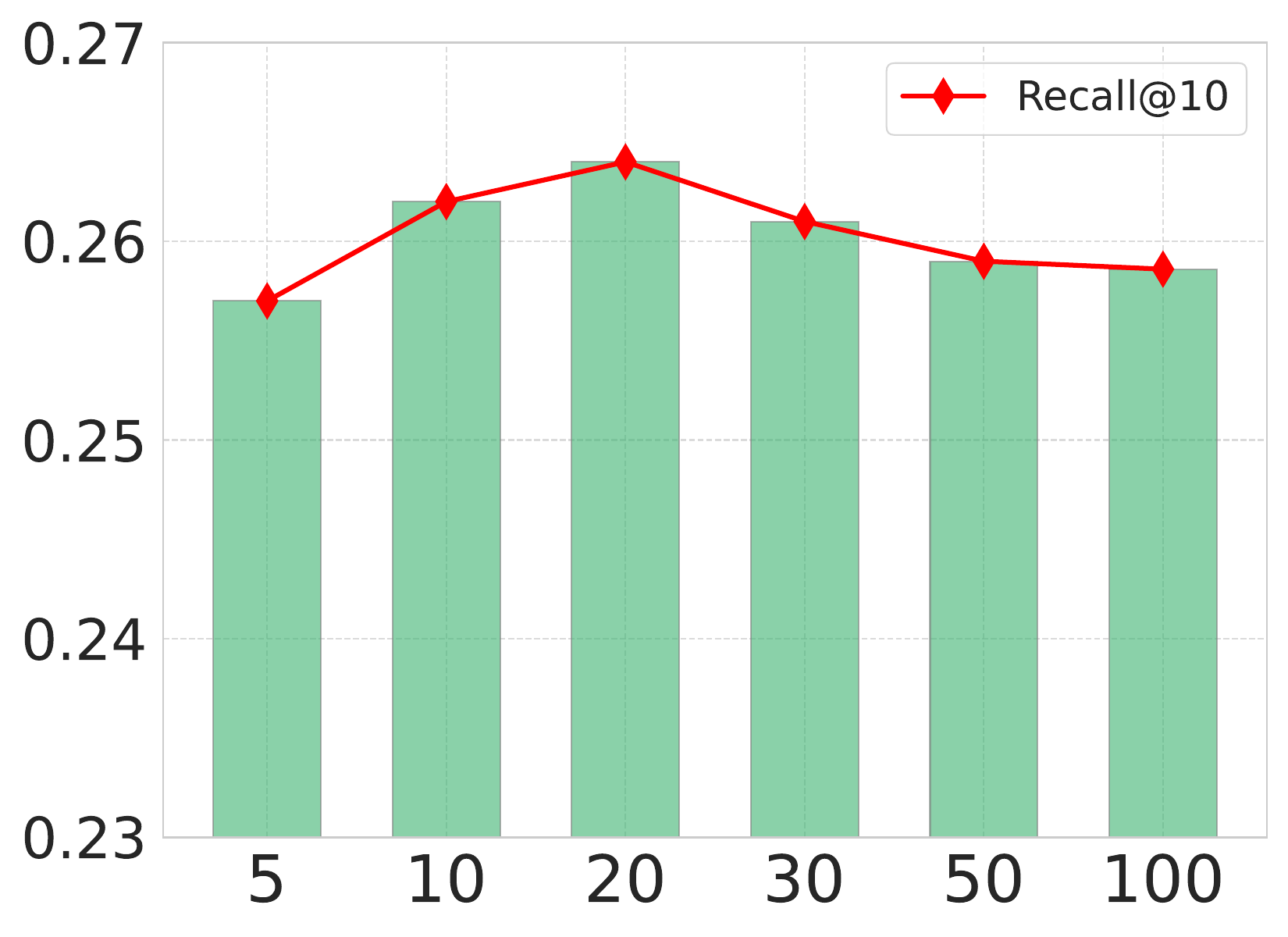}
        \caption*{(a) ReDial ($k$)}
    \end{minipage}\hfill
    \hfill
    \begin{minipage}[t]{0.23\textwidth}
        \centering
        \includegraphics[width=\textwidth]{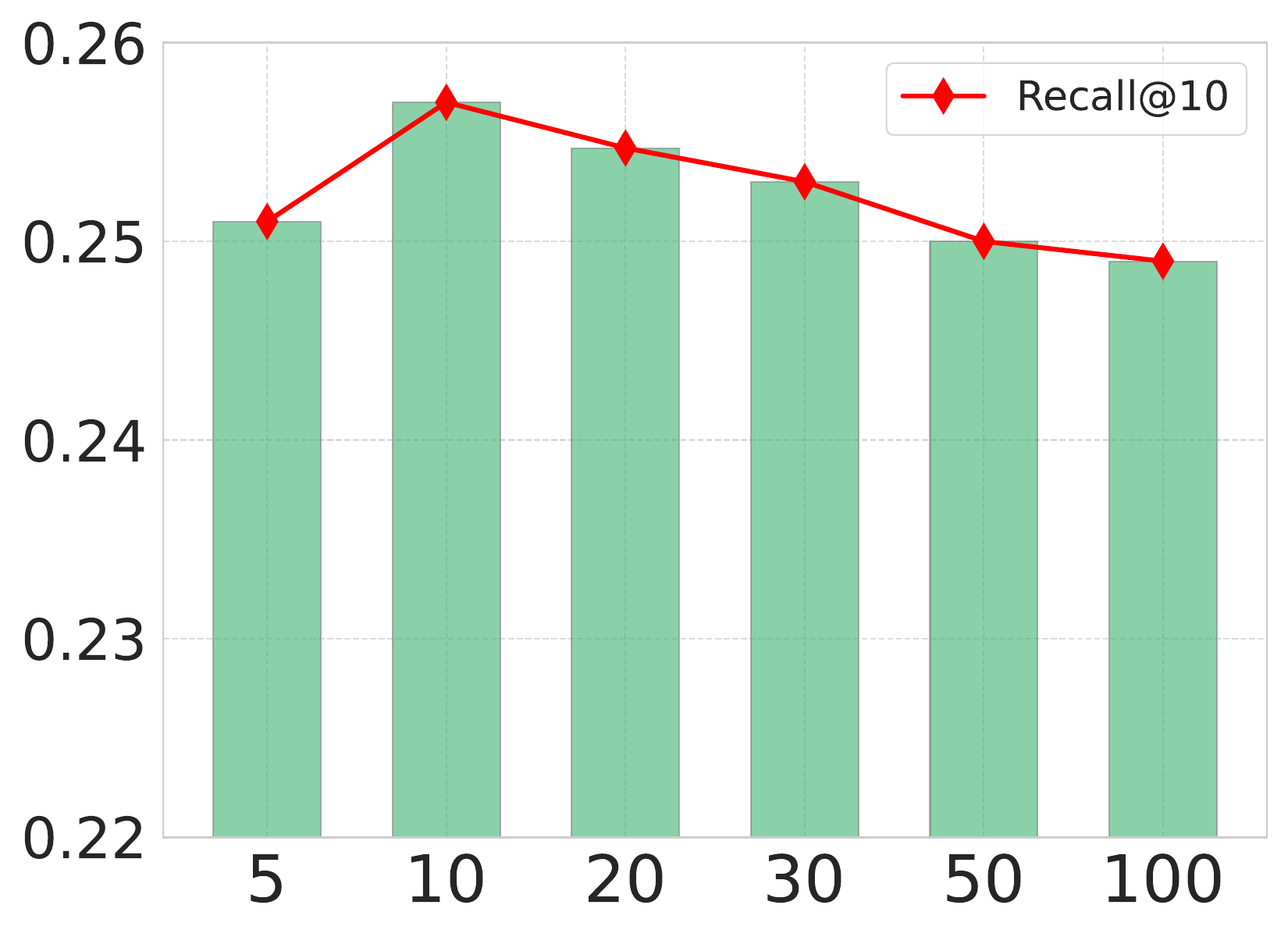}
        \caption*{(b) INSPIRED ($k$)}
    \end{minipage}
    
    \begin{minipage}[t]{0.23\textwidth}
        \centering
        \includegraphics[width=\textwidth]{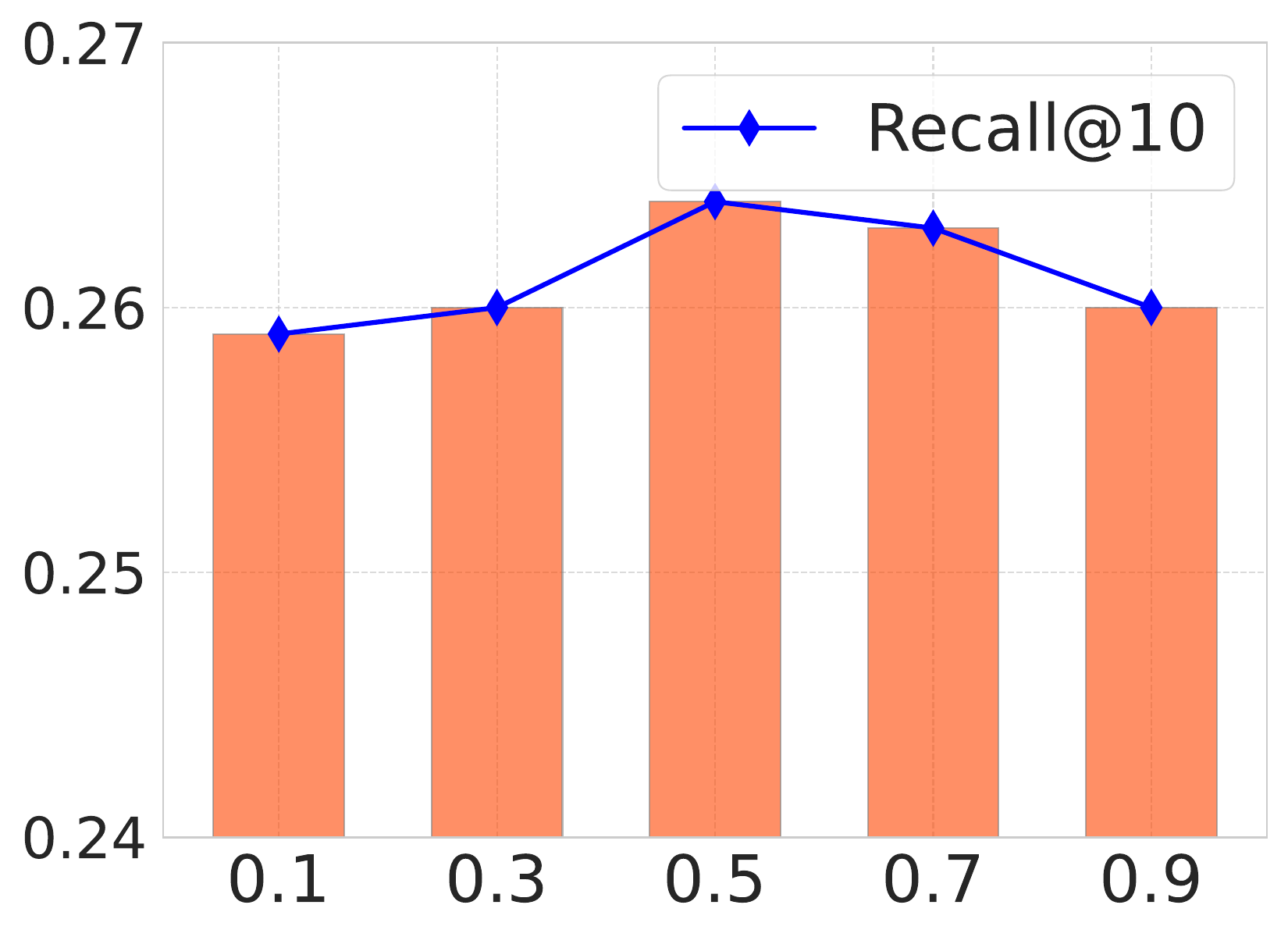}
        \caption*{(c) ReDial ($\lambda$)}
    \end{minipage}\hfill
    \hfill
    \begin{minipage}[t]{0.23\textwidth}
        \centering
        \includegraphics[width=\textwidth]{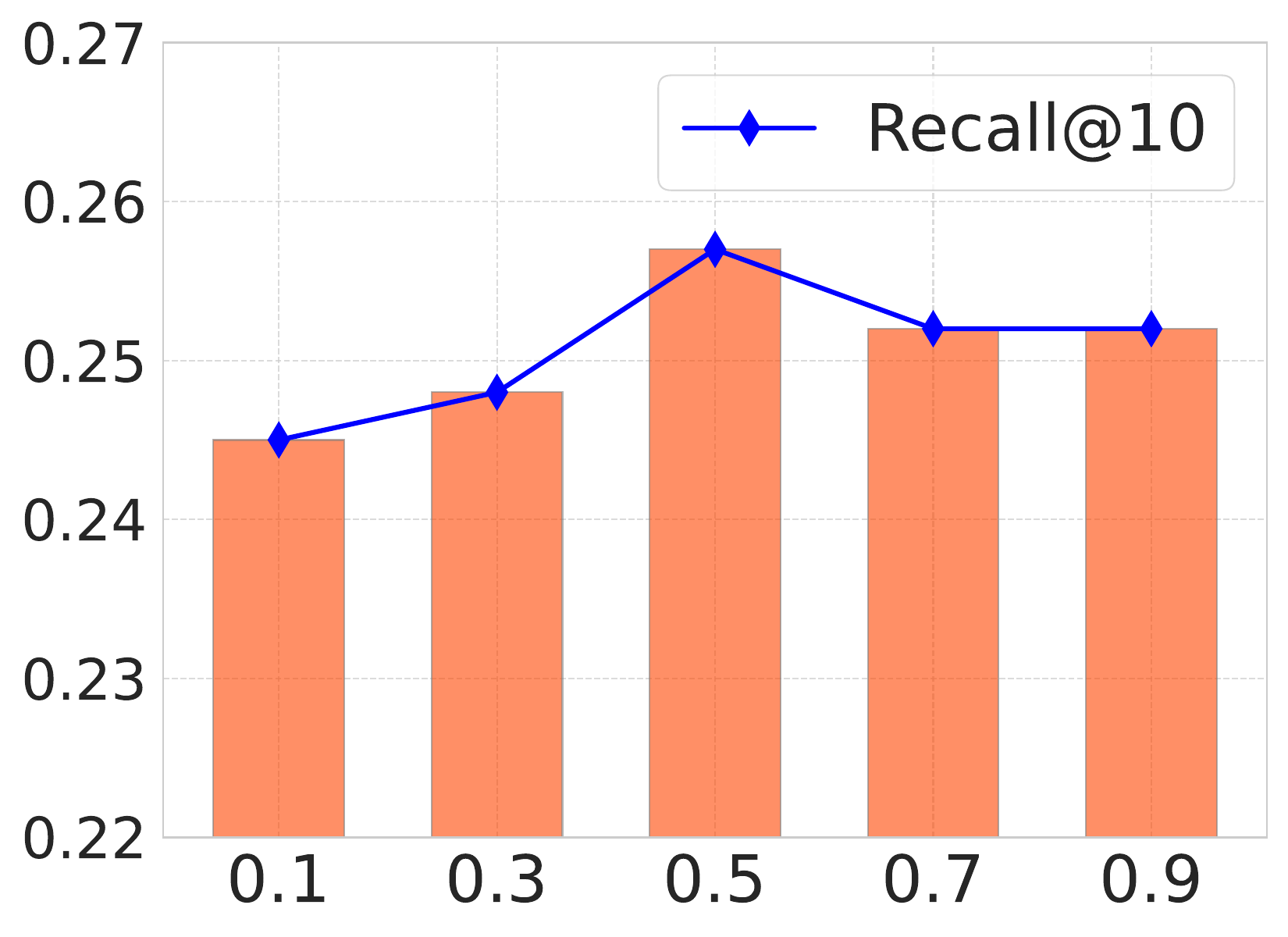}
        \caption*{(d) INSPIRED ($\lambda$)}
    \end{minipage}
    \caption{The impact of different $k$ and $\lambda$ values on Recall@10 for the ReDial and INSPIRED datasets.}
    \label{fig:combined_k_lambda}
\end{figure}

In our proposed image and textual semantic graphs, we keep the top $k$ items with the highest semantic relevance using the $k$-NN method \cite{knn}. We investigate the impact of different $k$ values (Eq. (\ref{eq:10})) on the model's performance. Specifically, we set the $k$ values to [5, 10, 20, 30, 50, 100]. From Figure \ref{fig:combined_k_lambda} (a) and (b), we can see that the ReDial and INSPIRED datasets exhibit similar trends in $k$ value variations. The Recall@10 for the ReDial dataset reaches a higher peak at \(k=20\), while the INSPIRED dataset achieves the best performance at \(k=10\). Overall, selecting an appropriate $k$ value is crucial. A small $k$ value may fail to capture enough relevant items, while a large $k$ value may introduce irrelevant noise, leading to a decline in performance. In our paper, we use the optimal $k$ values \(k=20\) on ReDial and \(k=10\) on INSPIRED in other experiments.

\subsubsection{Effect of $\lambda$}
The parameter \(\lambda\) controls the fusion ratio between the textual semantic graph and the image semantic graph. We investigate the impact of different fusion ratios of \(\lambda\) on model performance. We adjust the value of \(\lambda\) within \([0.1, 0.3, 0.5, 0.7, 0.9]\). From Figure \ref{fig:combined_k_lambda} (c) and (d), we observe that as \(\lambda\) varies, the Recall@10 metric exhibits similar trends on both the INSPIRED and ReDial datasets. On the ReDial dataset, the Recall@10 value peaks around \(\lambda = 0.5\) before slightly declining. Similarly, on the INSPIRED dataset, the Recall@10 value reaches its peak around \(\lambda = 0.5\) before starting to decline. These results indicate that the fusion ratio significantly influences model performance, and an optimal range of \(\lambda\) can balance the contributions of both textual and image modalities. We use the optimal values \(\lambda = 0.5\) on ReDial and \(\lambda = 0.5\) on INSPIRED in other experiments in our paper.

\section{CONLUSION AND FUTURE WORK}
In this paper, we propose MSCRS, a novel multi-modal semantic graph prompt learning framework for CRS. Our approach integrates textual, image, and collaborative semantic information by constructing three semantic graphs to enhance entity representations and user preference modeling. In addition, by incorporating prompt learning with GNN-based neighborhood aggregation, MSCRS provides an LLM with topological cues, effectively guiding it to extract relevant information from text inputs and generate high-quality responses. Extensive experiments on recommendation and conversational tasks demonstrate that MSCRS improves performance in both item recommendation and response generation. 

\begin{acks}
This research was supported by the National Natural Science Foundation of China (62402093) and, the Sichuan Science and Technology Program (2025ZNSFSC0479). This work was also supported in part by the National Natural Science Foundation of China under grants U20B2063 and 62220106008, and the Sichuan Science and Technology Program under Grant 2024NSFTD0034. 
\end{acks}
\clearpage
\bibliographystyle{ACM-Reference-Format}
\bibliography{sample-base}
\end{document}